\documentclass{v16nufact}

\def\Journal#1#2#3#4{{#1} {\bf #2}, #3 (#4)}

\def\be{\begin{equation}}
\def\ee{\end{equation}}
\def\bea{\begin{eqnarray}}
\def\eea{\end{eqnarray}}

\newcommand{\Photo}{}

\begin{document}
\vspace*{4cm}
\title{First Measurement of Neutrino Interactions in MicroBooNE}

\author{ P. A. Hamilton }

\address{Fermilab, Wilson Hall 10W, Batavia, Illinois 60510, USA}

\maketitle\abstracts{
The MicroBooNE detector has recently completed its first year of neutrino beam data-taking in the Booster Neutrino Beam at Fermilab, having collected approximately half of its intended data ($3.4\times10^{20}$ of $6.6\times10^{20}$ protons on target). We present kinematic 
distributions of neutrino interactions observed from a small subset of this data (equivalent to $5\times10^{19}$ protons on target), both as a first step towards a charged-current muon neutrino cross-section on argon, and as an exploration of 
the capabilities and operational challenges of large liquid argon time projection chambers as neutrino detectors. These distributions have been assessed using fully automated event selection and reconstruction.}

\section{Introduction}
Over the next decade, long-baseline neutrino oscillation experiments will move forwards into a ``precision era'' of oscillation measurements, with the proposed next-generation experiments Dune~\cite{dune} and Hyper-K~\cite{hk} at the forefront of this effort. For these experiments 
to reach their target sensitivities, systematic uncertainties must be reduced as well. Some of the most difficult systematic uncertainties on our current oscillation measurements arise from uncertainties in neutrino-nucleus cross-sections. This is 
particularly relevant for DUNE, since the world's cross-section knowledge for its target nucleus (argon) in the intended energy range of the DUNE beam is extremely scarce~\cite{agnt1}~\cite{agnt2}.

MicroBooNE is positioned to greatly expand our knowledge of neutrino-argon cross-sections, as well as its primary goal of probing the MiniBooNE low-energy excess~\cite{mnbn}. As part of the Fermilab short-baseline programme, it offers a large argon target in an intense neutrino beam, giving it the capacity for high-statistics 
cross-section measurements of muon neutrinos with energies from a few hundred MeV up to 2 GeV. As a liquid argon time projection chamber (TPC), it also offers excellent reconstruction and distinction between final state topologies, giving it the capacity to measure cross-sections of 
highly specific interaction channels, and thus to probe tensions in existing nuclear models.

Many cross-section analyses are currently underway within the MicroBooNE collaboration, including studies of proton multiplicity, kaon production, neutral-current channels, and more. Presented here are kinematic distributions from the first and most 
developed MicroBooNE cross-section analysis, the charged-current inclusive cross-section. This analysis will provide a foundation for comparison to other experiments, and for the development of the tools required for more specific cross-section 
analyses in future.

\section{The MicroBooNE Detector}
\begin{figure}[h!]
\centerline{\includegraphics[width=0.4\linewidth]{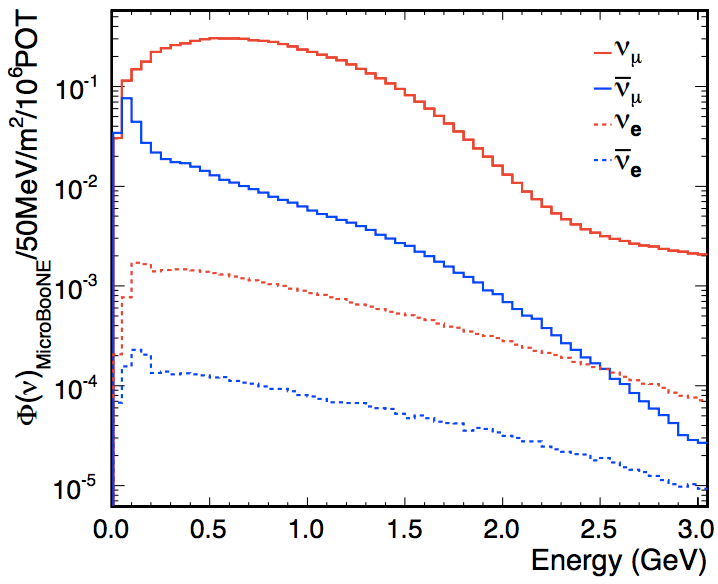}}
\caption[]{The BNB energy spectrum for neutrinos and antineutrinos of both muon and electron flavours.}
\label{fig:BNBspec}
\end{figure}

The MicroBooNE detector is situated in the Booster Neutrino Beam (BNB) at Fermilab, receiving a broad energy spectrum of muon neutrinos (shown in Figure \ref{fig:BNBspec}). MicroBooNE also samples the NuMI beam at an off-axis angle of 135 mrad, allowing comparison with neutrinos of a markedly different energy spectrum. Only BNB data has been used 
for this analysis.

MicroBooNE itself is a liquid argon TPC with an active mass of 89 tons --- the largest liquid argon TPC yet operated in the USA, with dimensions of $10.4 \times 2.5 \times 2.3$ m. The TPC is instrumented with three charge readout wire planes (each with a 3 mm spacing between wires): the vertical collection plane, and two induction 
planes at $\pm 60^{\circ}$. The scintillation light from interactions in the argon is read out by 32 8'' cryogenic PMTs. The operating principles of this detector can be read in the MicroBooNE technical design report \cite{tdr}.

\begin{figure}[h!]
\begin{minipage}{0.33\linewidth}
\centerline{\includegraphics[width=0.9\linewidth]{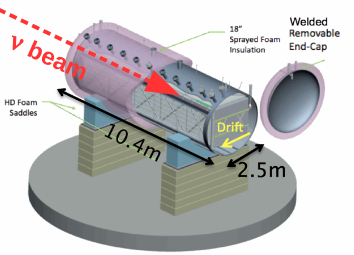}}
\end{minipage}
\hfill
\begin{minipage}{0.32\linewidth}
\centerline{\includegraphics[width=0.9\linewidth]{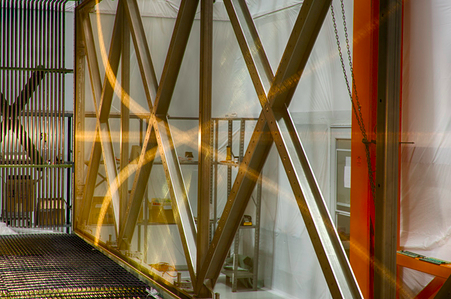}}
\end{minipage}
\hfill
\begin{minipage}{0.32\linewidth}
\centerline{\includegraphics[width=0.9\linewidth]{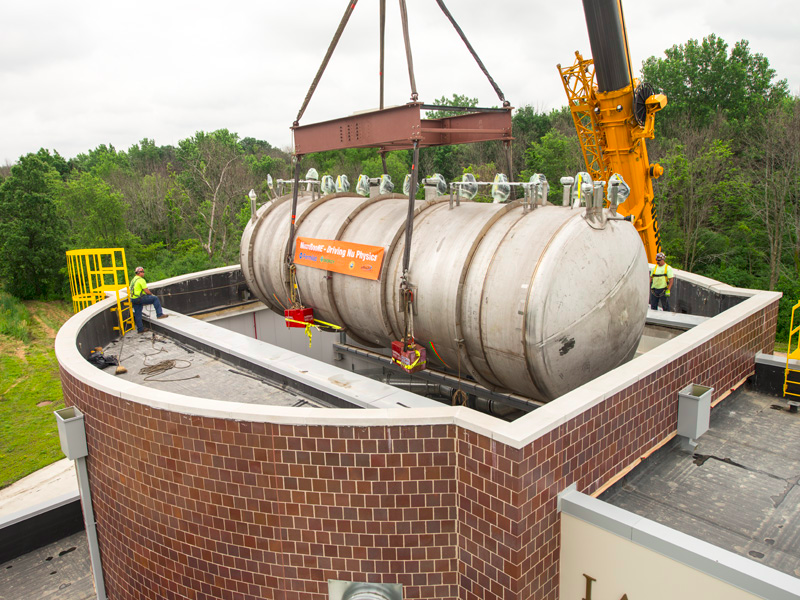}}
\end{minipage}
\caption[]{A schematic of the MicroBooNE detector showing its dimensions with respect to the beam direction (left); a photograph of the wire planes before installation (centre), and a photograph of the assembled TPC plus cryostat being lowered into the Liquid Argon Test Facility (LArTF) (right).}
\end{figure}

\begin{figure}[h!]
\centerline{\includegraphics[width=0.85\linewidth]{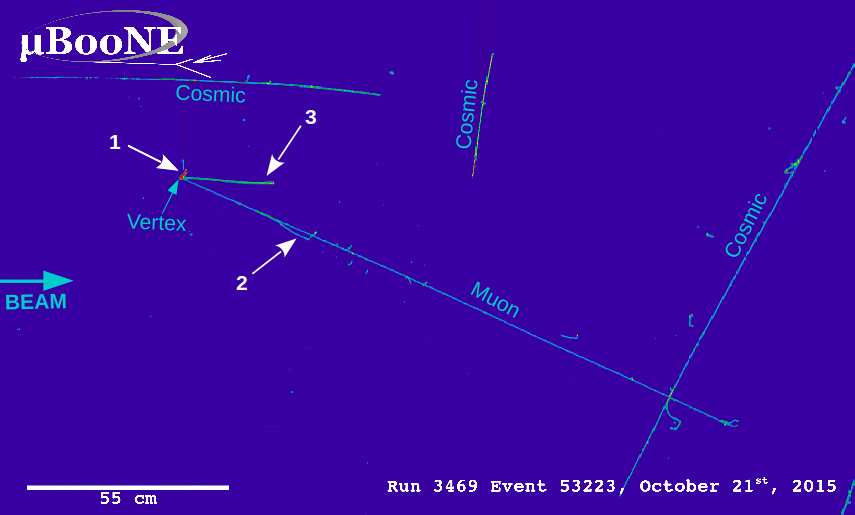}}
\caption[]{An annotated event display from MicroBooNE data. The colour scale denotes received charge. This event shows a beam neutrino interaction and demonstrates: \\
(1) The capacity to identify low-energy nuclear ejecta, as shown by the highly ionising (red) proton track stub. \\
(2) The capacity for high-resolution tracking, as shown by the clearly distinguished delta radiating from the longest track (presumed to be the muon).\\
(3) The capacity for step-by-step calorimetry, as shown by by the gradient in charge towards the Bragg peak at the end of the second longest track. 
}
\label{fig:evedisp}
\end{figure}

This detector setup allows for MicroBooNE to perform extremely detailed reconstruction of neutrino interactions. An example event display is shown in Figure \ref{fig:evedisp}, exhibiting features that demonstrate some of these advantages.

\section{The MicroBooNE Charged-Current Inclusive Selection}

In order to make high-statistics physics measurements, MicroBooNE requires automated selection algorithms, which in turn require automated reconstruction as input. This is provided by the LArSoft software package~\cite{larsoft}. LArSoft is a suite of tools designed for 
use in liquid argon TPCs across the US neutrino programme, providing 3D hit-finding, tracking, particle identification, electromagnetic shower reconstruction and optical reconstruction --- all the tools required for a complete picture of neutrino interactions in a liquid argon TPC.

MicroBooNE uses the output of LArSoft in two complementary (preliminary) selections to select charged-current interactions in the liquid argon. Both are fully automated, cut-based selections. The results presented here are those given by the second selection, but the full details of both selections can be read in MicroBooNE public note 1010~\cite{uboone_1010}.

\begin{figure}[h!]
\centerline{\includegraphics[width=0.6\textwidth]{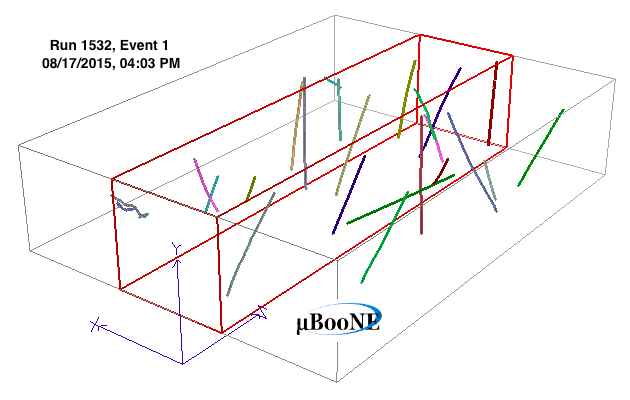}}
\caption{This event display shows real data from MicroBooNE. The drift window of 1.6 ms after the trigger is outlined in red; the grey boxes to either side are the pre-drift and post-drift readout windows (each of equal duration to the trigger window). 
The figure shows that in the time taken to read out a single beam event the detector samples tens of cosmic tracks, represented here by coloured lines. Each line is a fully reconstructed track, showing that MicroBooNE's reconstruction is already 
performing high-fidelity tracking of cosmic rays.}
\label{fig:cosmics_cosmics_everywhere}
\end{figure}

The largest background for charged-current neutrino interactions in MicroBooNE comes from cosmic rays. Situated as it is at ground level, the detector sees tens of cosmic rays passing through it per drift window (an example of this being shown 
in Figure \ref{fig:cosmics_cosmics_everywhere}). 

The cosmic backgrounds are removed in four stages:

\begin{itemize}
 \item An initial cut looking for a light signal above 50 P.E. in the beam window, indicating the possible presence of a beam neutrino interaction.
 \item A ``flash matching'' cut, which looks for a multi-PMT coincident light signal (above 50 P.E.) within 70 cm of the candidate muon track. This identifies beam events by looking for their characteristic scintillation light profile.
 \item A fiducial volume cut, which rejects vertices too close to the detector walls (20 cm in the vertical dimension, 10 cm in the others). This removes the majority of entering backgrounds.
 \item A branching set of topology-based cuts for events with charged particle multiplicity 1, 2 and $> 2$. These allow for enhanced efficiency in high-multiplicity events (where the population of cosmic backgrounds is low), while maintaining good 
 rejection at low multiplicities, and enabling the targeted rejection of Michel decays from stopping cosmic muons in two-track events.
\end{itemize}

\begin{figure}[h!]
\centerline{\includegraphics[width=0.49\textwidth]{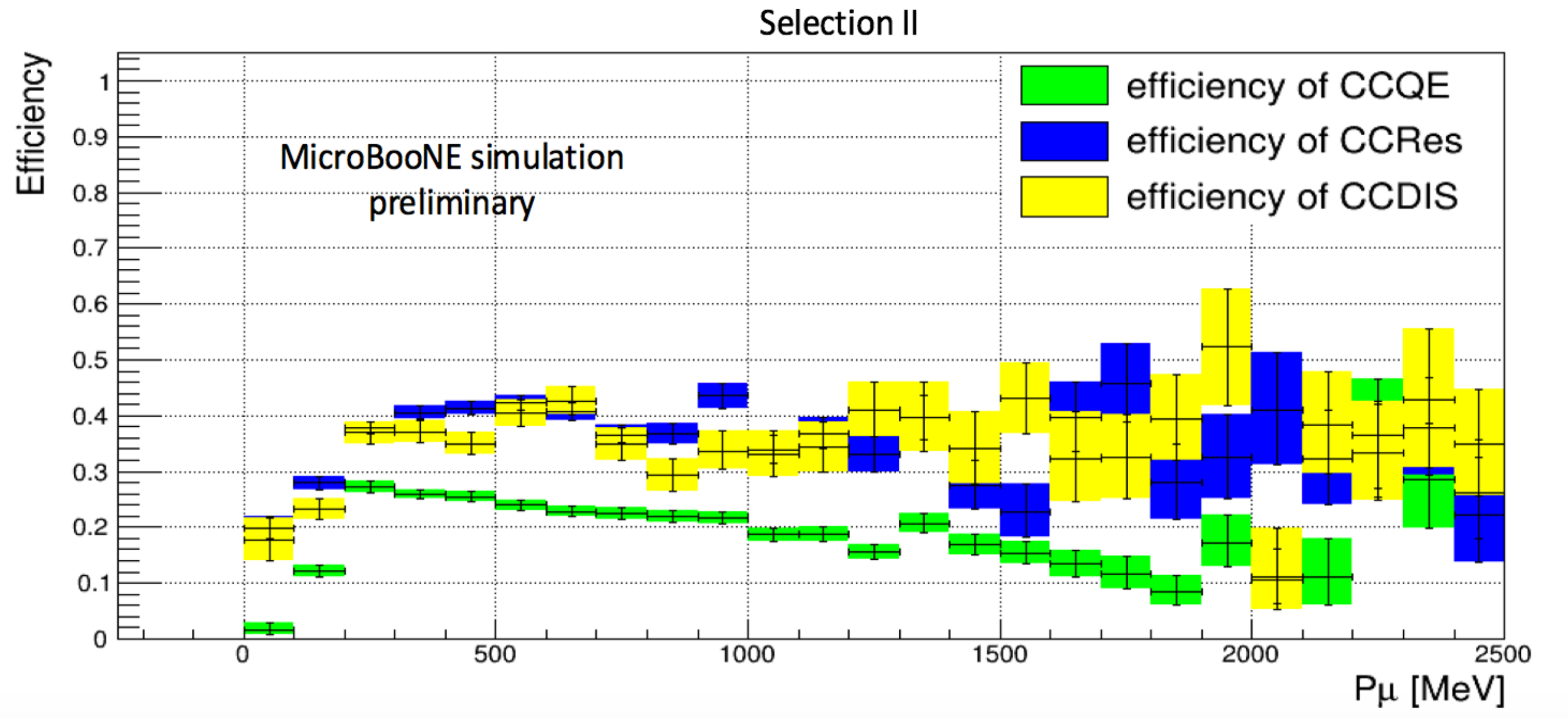}}
\centerline{\includegraphics[width=0.49\textwidth]{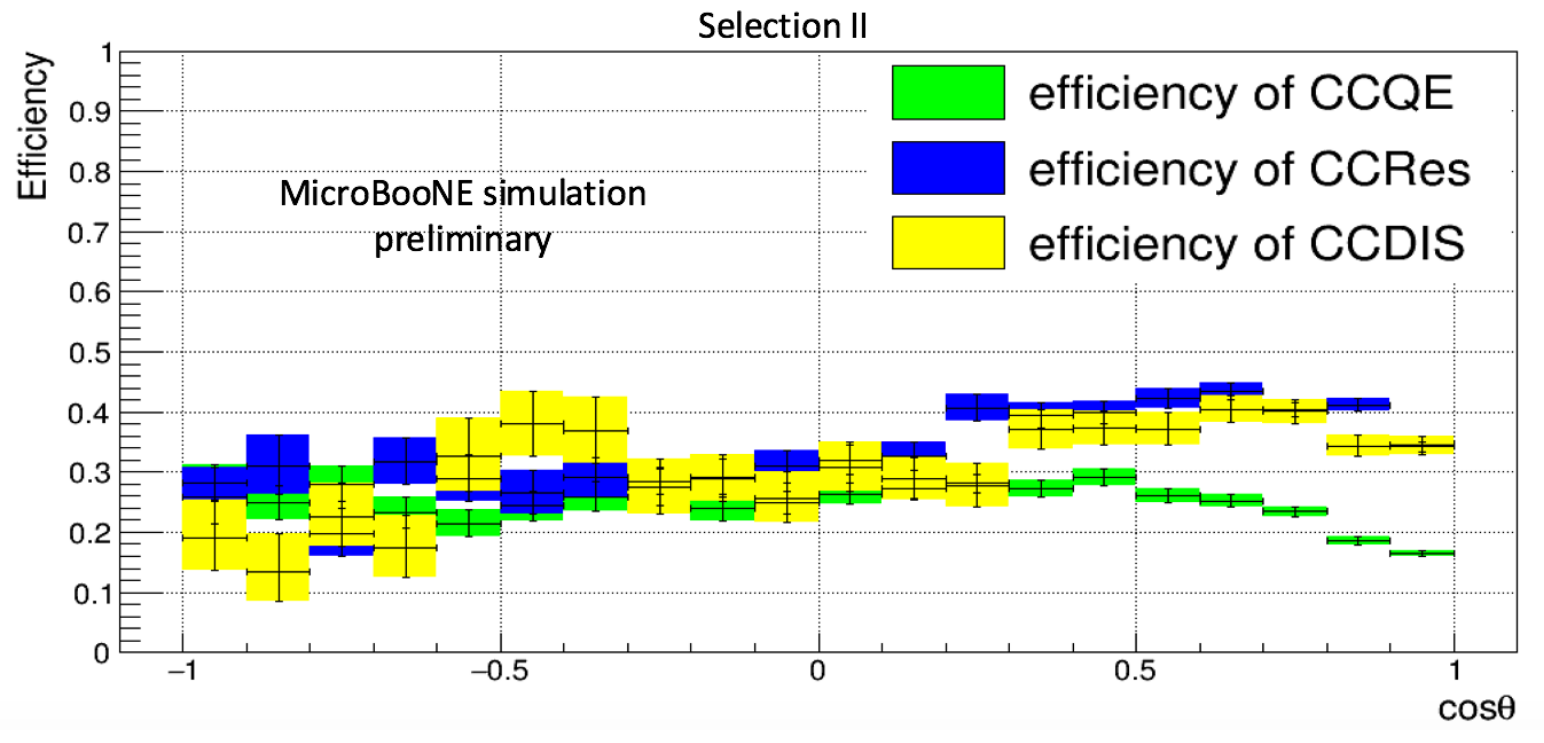}
\includegraphics[width=0.49\textwidth]{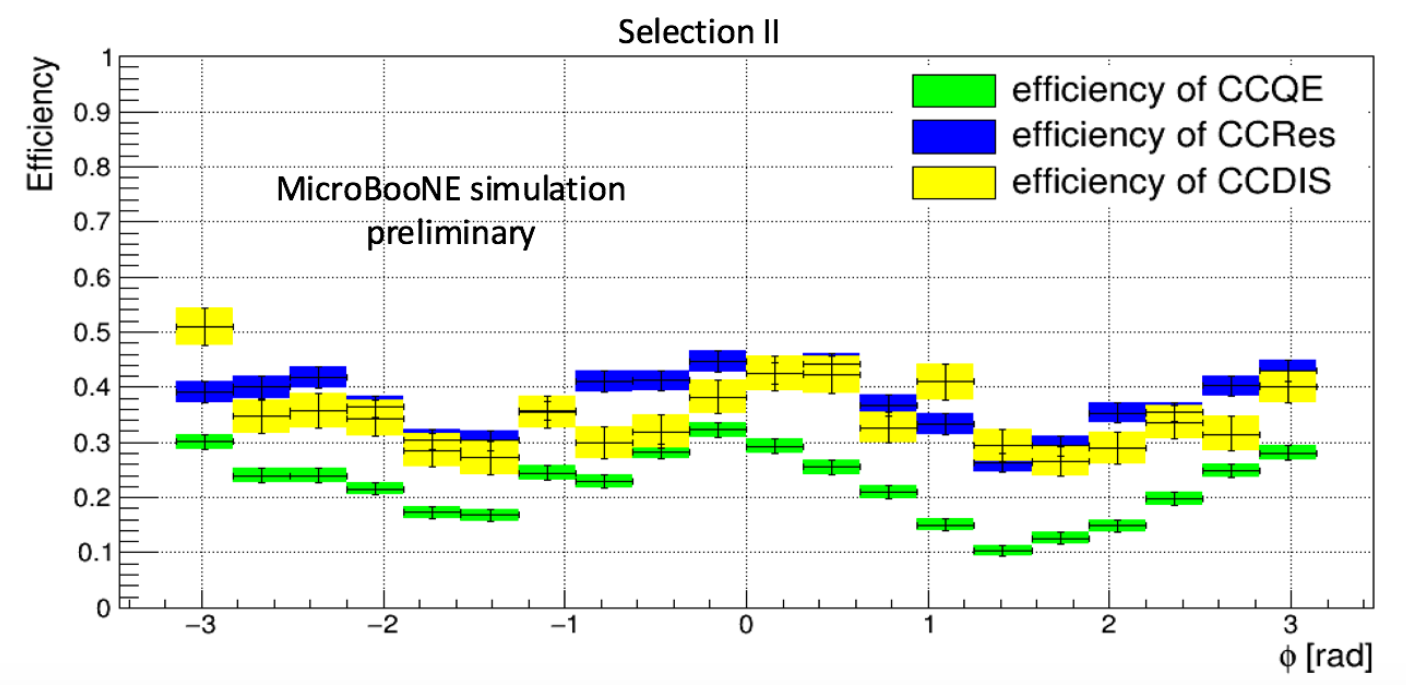}}
\caption[]{These plots show the selection efficiency as a function of muon momentum (top), angle with respect to the beam direction (bottom left), and azimuthal angle around the beam axis (bottom right). Troughs in the efficiency at $\phi = \pm \frac{\pi}{2}$ show the 
suppression of signal from cuts targeting predominantly vertical cosmic backgrounds. The decline in efficiency for high-energy CCQE events arises from a containment requirement on events with only a single track. Note that the efficiency losses shown here include the 
losses due to detector acceptance.}
\label{fig:eff}
\end{figure}

These cuts give an overall efficiency $\times$ acceptance of 30\%, and an overall purity of 65\%. The efficiency is shown in Figure \ref{fig:eff} as functions of the muon momentum, the muon track's angle with respect to the beam direction ($\theta$), and the muon track's azimuthal angle around the beam axis ($\phi$).

\begin{figure}[h!]
\centerline{\includegraphics[height=0.2\textheight]{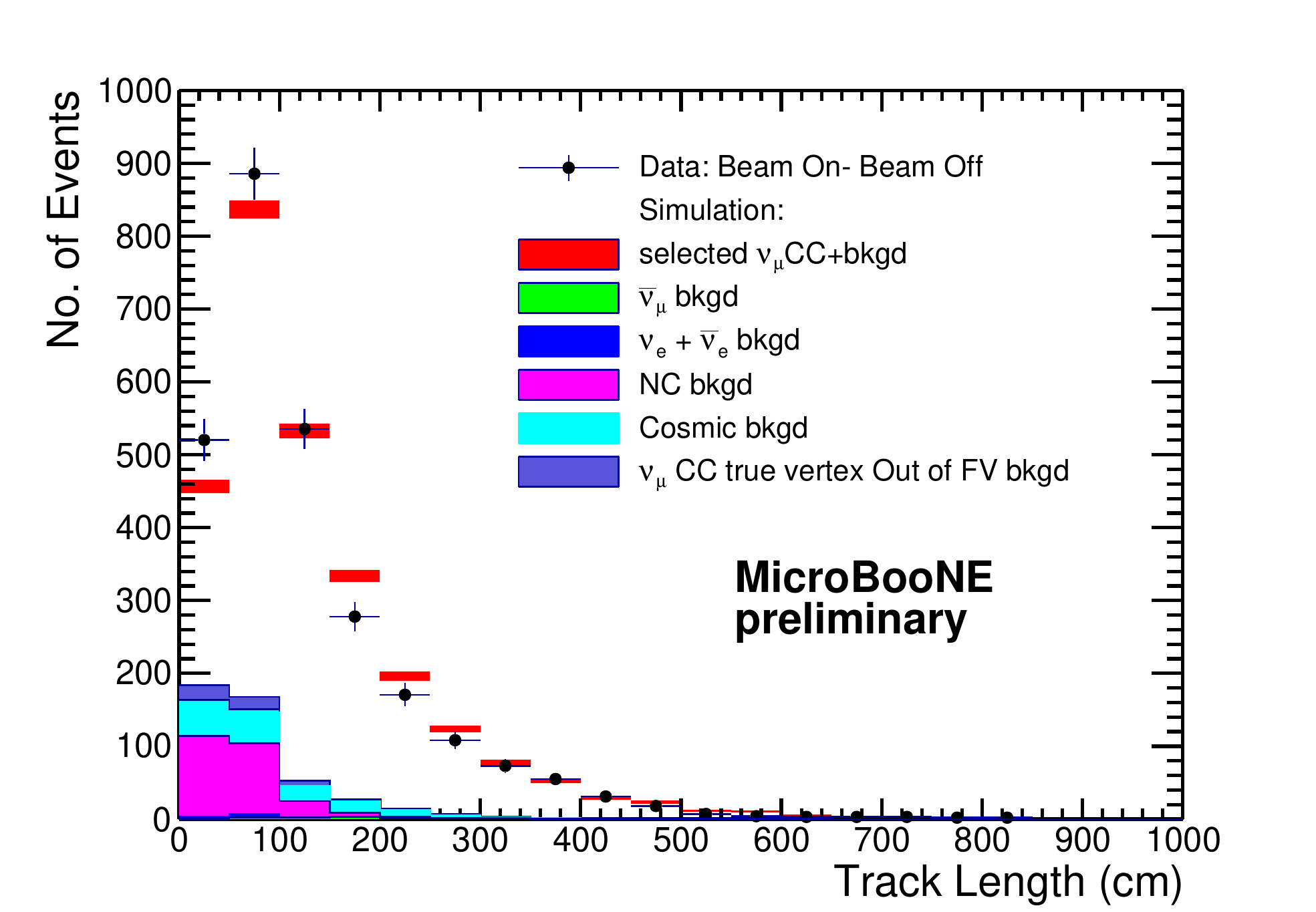}}
\centerline{\includegraphics[height=0.2\textheight]{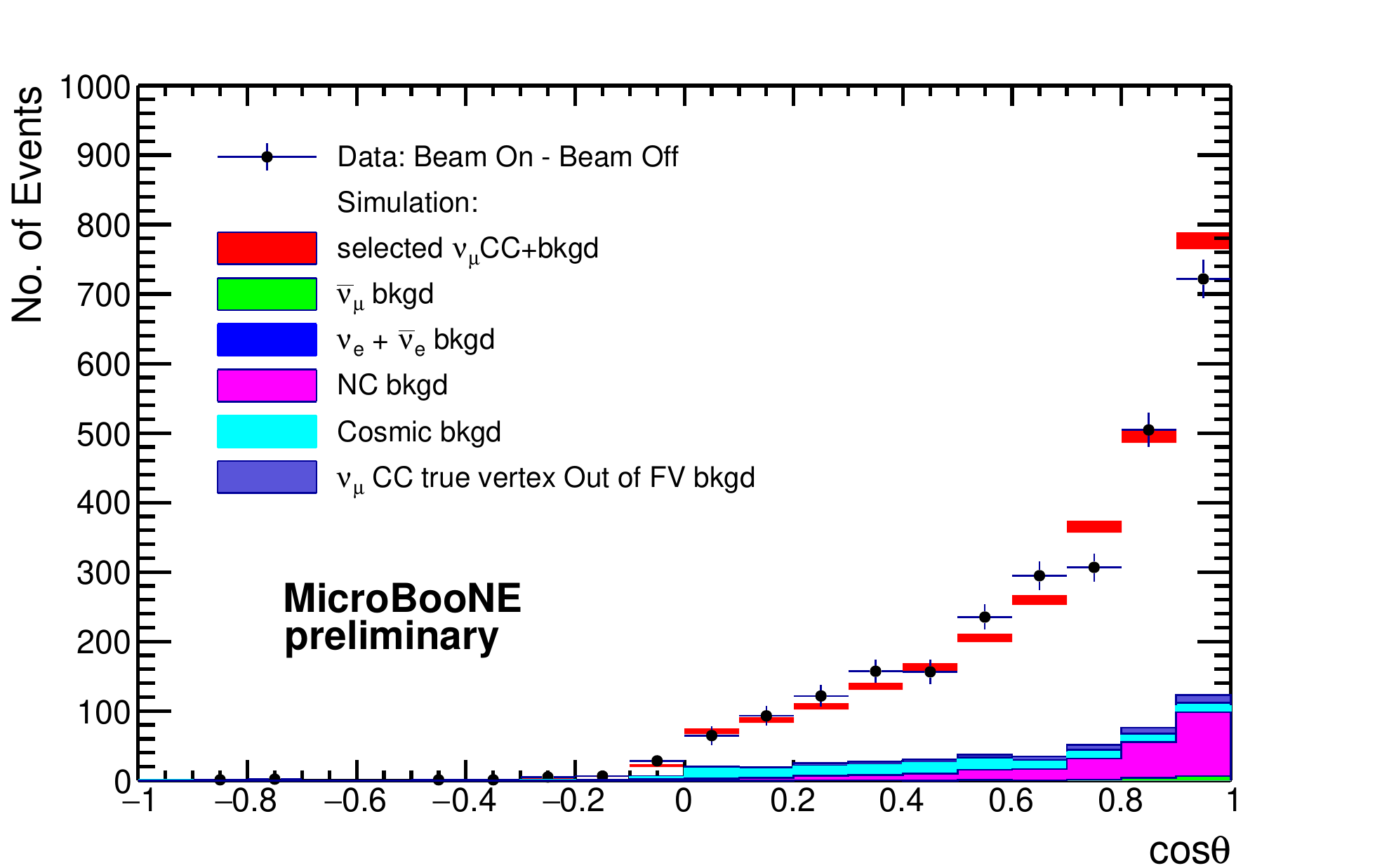}
\includegraphics[height=0.2\textheight]{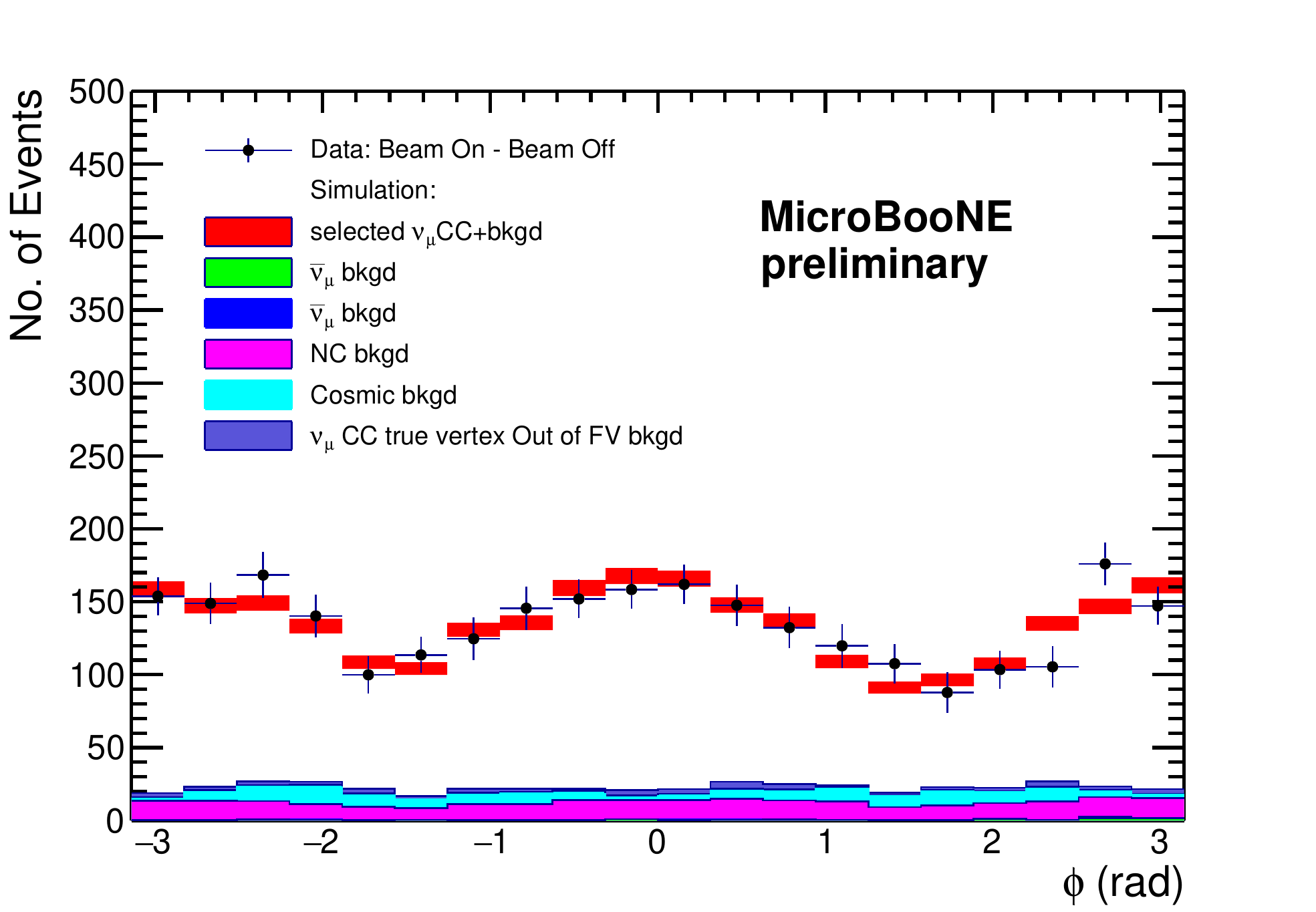}}
\caption[]{These plots show the kinematic distributions produced by the selection, in track length (top), angle with respect to the beam direction (bottom left), and azimuthal angle around the beam axis (bottom right). Agreement between data and Monte Carlo
appears good. The distribution in $\phi$ shows the impact of our selection in shaping the accepted signal as discussed in Figure \ref{fig:eff}; future improvements to cosmic tagging will allow us to relax the harsh cuts producing this shape. The off-beam backgrounds 
have been subtracted using estimates from data.}
\label{fig:results}
\end{figure}

Applying this selection to a subset of our data equivalent to $5\times10^{19}$ protons on target, and performing a background subtraction of the surviving cosmics outside the beam window using beam-off data, we observe the kinematic distributions shown in Figure \ref{fig:results}. These distributions are presented with statistical uncertainties only, and with Monte Carlo distributions scaled to the same number of events as the data; the assessment of the systematic 
uncertainties is still ongoing. The key systematics to be assessed in converting future kinematic distributions into differential cross-sections are:
\begin{itemize}
 \item The uncertainty on the BNB flux.
 \item Detector effects, of which the most significant are the liquid argon purity and the effect of distortions in the electric field (also known as space charge effects).
 \item Model uncertainties in our simulation of neutrino-nucleus interactions.
\end{itemize}
The flux uncertainty is known to be approximately 10\%~\cite{flux}. The systematic error from model uncertainties is being assessed by a programme of reweighting the GENIE~\cite{genie} model input parameters. The systematic errors from detector effects are the subject of 
studies on MicroBooNE data.

\section{Future Work}

The work done to produce the kinematic distributions shown will be the basis for future cross-section measurements in each of the relevant kinematic variables. For these final measurements our assessment of the surviving cosmic background will be 
enhanced by the implementation of cosmic overlays --- taking tracks from cosmic data and superimposing them on beam Monte Carlo, rather than using simulated cosmic tracks. We will also complete our assessment of the systematic uncertainties 
associated with this measurement, and expand our statistics from the $5\times10^{19}$ protons on target used here (with MicroBooNE being scheduled to collect $6.6\times10^{20}$ protons on target over its full lifetime).

This analysis represents only the first step in the MicroBooNE cross-section programme and has been instrumental in developing the tools to carry out more topology-specific cross-section analyses (e.g. pion-less charged current, neutral current elastic). 
The potential of liquid argon TPC technology to enable measurements of the cross-section as a function of the outgoing charged particle multiplicity down to very low particle energies --- and to identify those particles with a high degree of accuracy 
--- is of particular interest. Improved software tools are being developed to carry these analyses forward, including improved shower reconstruction, improved tracking over regions with unresponsive wire channels (``dead wires''), improved matching between TPC tracks and their associated 
scintillation light, and selection tools based on deep learning.

The challenges faced in this selection in overcoming the cosmic background will also be addressed through a physical upgrade to the detector. Over the 2016 summer shutdown, a cosmic ray tagger system will be installed in the LArTF facility housing 
MicroBooNE to tag cosmic tracks entering the detector from above. This should provide significant additional power to reject cosmic backgrounds without incurring additional losses in signal efficiency.

\section{Conclusion}

MicroBooNE has made its first kinematic measurements of neutrino interactions in a muon neutrino beam. A charged-current inclusive cross-section will soon follow. These measurements demonstrate that a large-scale liquid argon TPC can make excellent measurements of 
neutrino interactions, and provide the foundation for a much more wide-ranging programme of cross-section measurements to come. The experiment's progress to date has been documented extensively through its public notes~\cite{uboone_1010}.

\end{document}